# Observation of band narrowing and mode conversion in two-dimensional binary magnonic crystal

Nikita Porwal[1], Koustuv Dutta[2], Sucheta Mondal[2], Samiran Choudhury[2], Jaivardhan Sinha[2], Anjan Barman[2,*] and P. K. Datta[1,*]

[1]*Department of Physics, Indian Institute of Technology Kharagpur, W.B. 721302, India*
[2]*Department of Condensed Matter Physics and Material Sciences, S. N. Bose National Centre for Basic Sciences, Block JD, Sector III,
Salt Lake, Kolkata 700 106, India*
[*]Email: abarman@bose.res.in, pkdatta.iitkgp@gmail.com

We introduce a new type of binary magnonic crystal, where $Ni_{80}Fe_{20}$ nanodots of two different sizes are diagonally connected forming a unit and those units are arranged in a square lattice. The magnetization dynamics of the sample is measured by using time-resolved magneto-optical Kerr effect microscope with varying magnitude and in-plane orientation ($\phi$) of the bias magnetic field. Interestingly, at $\phi = 0°$, the spin-wave mode profiles show frequency selective spatial localization of spin-wave power within the array. With the variation of $\phi$ in the range $0°<\phi\leq45°$, we observe band narrowing due to localized to extended spin-wave mode conversion. Upon further increase of $\phi$, the spin-wave modes slowly lose the extended nature and become fully localized again at 90°. We have extensively demonstrated the role of magnetostatic stray field distribution on the rotational symmetries obtained for the spin-wave modes. From micromagnetic simulations, we find that the dipole-exchange coupling between the nano-dots leads to remarkable modifications of the spin-wave mode profiles when compared with arrays of individual small and large dots. Numerically, we also show that the physical connection between the nano-dots provides more control points over the spin-wave propagation in the lattice at different orientations of bias magnetic field. This new type of binary magnonic crystal may find potential applications in magnonic devices such as spin-wave waveguide, filter, coupler, and other on-chip microwave communication devices.

# I. Introduction:

Nanomagnets have huge applications in magnetic storage [1], memory [2], logic [3], sensors [4], other spintronics and biomedical devices [5, 6]. One of the more recent and emerging fields based on nanomagnetism is magnonic crystals (MCs) [7], which are periodically modulated ferromagnetic materials, such as ferromagnetic nanodots [8], nanowire [9], nanoscale antidot arrays [10], where spin waves (SWs) are carrier waves. Due to their wavelength falling in the nanoscale regime for GHz to sub-THz frequency range spin-waves, MCs are ideally suited for nanoscale on-chip microwave communication devices. Those are also capable of forming magnonic minibands with allowed and forbidden frequencies [11-13]. By varying the physical and geometrical parameters of the artificial crystal, the nature of intra-element and inter-element magnetic field distributions within the array can be tuned, which in turn, modify its SW dynamics. A plethora of studies on the quasi-static and dynamic properties of 1-D, 2-D and 3-D MCs have been carried out due to their fundamental physical properties and their promising applications, such as spin-wave filter, coupler, phase shifter, splitter and other magnonic devices [14-23]. Recently, the entanglement between various spin-based phenomena has emerged as a new field, coined as magnon-spintronics [24].

Analogous to various natural or artificial crystals, introduction of bi- or multi-components can lead to a variation in the periodic potential for the propagating SWs offering large tunability in SW spectra, band structure, band gap and propagation velocities. This fueled an upsurge of research interest in bi-component and binary magnonic crystals during the last one decade. This ranges from filled antidot systems [2501], both in physically connected and isolated forms, nanowires [31-32] or nanodots [33] of different materials placed in close proximity, to nanodots [34] of same material but of different size or shape forming the basis structure of the crystal. The fabrication of bi-component magnonic crystal needs multistep lithography, which is non-trivial when the two components have physical connection [35], but they show distinct advantages such as significant improvement of SW propagation velocity. Such advantages may be readily derived from physically connected binary magnonic crystal (BMC) with much easier fabrication process. However, little efforts have been made to this end so far. This motivates us to fabricate and study the SW dynamics in a new type of BMC with nanodots of two different sizes connected at one corner and dispersed over a square lattice. The SW dynamics of this BMC is studied using time-resolved magneto-optical Kerr effect (TR-MOKE) microscope by varying the magnitude and in-plane orientation ($\phi$) of the bias magnetic field. At $\phi = 0°$, SW modes show frequency selective spatial localization of SW power within the array. Appearance of two separate bands are observed in the SW spectra. With the variation of $\phi$ in the range $0°<\phi\leq45°$, the lower frequency band gets significantly blue shifted. This leads to narrowing of the entire band due to conversion of SW modes from localized to extended nature. Upon further increase of $\phi$, the SW modes slowly become fully localized again at 90°. The role of magnetostatic field distributions on the rotational symmetries of the SW modes is demonstrated from the micromagnetic analysis. We have also compared the SW dynamics of BMC with individual arrays of large and small dots by simulation. Interestingly, physical connection between the nano-dots in the BMC is found to be significant for the SW propagation in this BMC at different orientations of the bias magnetic field.

## II. Sample Fabrication:

The sample is prepared by a combination of both electron-beam evaporation (EBE) and focused ion beam (FIB) lithography. Here, $Ni_{80}Fe_{20}$ (permalloy, Py hereafter) square nanodots of 15 nm thickness and two different sizes; small nanodots of 150 nm width and large nanodots of 300 nm width, are diagonally connected at one corner and arranged on a square lattice with lattice constant a = 450 nm. A 15-nm-thick Py film is first deposited on top of a self-oxidized Si (100) substrate using electron-beam evaporation in an ultrahigh vacuum chamber at a base pressure of $2 \times 10^{-8}$ Torr. The film is then immediately transferred to a sputtering chamber for the deposition of a capping layer of 5-nm-thick $SiO_2$ (by rf sputtering) on top of the Py film at a base pressure of $2 \times 10^{-7}$ Torr and Ar pressure of $5 \times 10^{-3}$ Torr to avoid degradation from the natural oxidation and exposure to high power laser during optical pump-probe experiment in normal environment. Proceeding to the next step, the binary magnonic crystal (BMC) is fabricated on the blanket Py film by using liquid $Ga^+$ ion beam lithography (Auriga-Zeiss FIB-SEM microscopes). The optimal values of voltage and current used for milling are 30 keV and 5 pA respectively to achieve A spot size of the beam of 50 nm. A scanning electron microscope (SEM) is used to confirm the actual sizes of the nanostructures. Figure 1(a) shows the SEM image of the samples with the nanodots arranged well-ordered in a square lattice. The size of the small dot is found to be ~150 ± 5 nm and that of large dot is found to be ~300 ± 5 nm.

## III. Experimental Details:

The TR-MOKE microscope used in our investigation is based upon a two-colour collinear optical pump-probe geometry [36]. Here, the second harmonic ($\lambda_{pump}$= 400 nm, repetition rate = 80 MHz, fluence = 10 mJ/cm$^2$, pulse width = 100 fs) of the fundamental beam of a mode locked Ti-Sapphire laser (Tsunami, Spectra Physics) is used as a pump the sample. A part of the fundamental beam ($\lambda_{probe}$= 800 nm, fluence = 2 mJ/cm$^2$, pulse width = 80 fs) is used to probe the time varying polar Kerr rotation from the sample. A delay stage situated at the probe path is used to create the necessary time delay between the pump-beam and the probe-beam with temporal resolution of about 100 fs limited by the cross-correlation of the pump and the probe beams. Finally, both the beams are combined together and focused at the centre of array by a microscope objective (N.A. = 0.65) in a collinear geometry. The probe-beam of spot size of about 800 nm diameter is tightly focused and overlapped with the slightly defocused pump-beam having larger diameter (~1 μm), at the centre of the array. Under this condition the probe collects the dynamics from the uniformly excited part of the sample. The sample is placed in a controllable piezo-electric scanning stage with high precession in x-y-z travel to place the pump-probe spots on the same position after rotating the sample to the desired angle. A static magnetic field is applied at a small angle (~10°) to the sample plane, the in-plane component of which is defined as the bias field *H*. The magnitude of the in-plane component of this field is chosen be large enough to saturate the magnetization within the sample plane. The time varying polar Kerr rotation and reflectivity are measured simultaneously at room temperature by using an optical bridge detector and and two separate lock-in amplifiers in a phase sensitive manner. The pump beam is modulated at 2 kHz frequency which eventually is being used as the reference frequency of the lock-in amplifier. Using this technique Kerr rotation and reflectivity signals are measured simultaneously avoiding any leakage of one into another. The measurement time window of about 2 ns used in this experiment is determined by the number of scan points and the time delay between pump and probe for two consecutive data points. Nevertheless, this 2 ns time window is found to be sufficient to resolve the SW frequencies for this sample.

**IV. Micromagnetic Simulation Details**:

Interpretation of the experimental results is aided by using micromagnetic simulations using OOMMF [37] here. In simulation, we apply a pulsed magnetic field of 30 Oe field strength and 10 ps rise/fall time, which reproduces the experimental conditions successfully. The method of the simulation can be found elsewhere [30]. The simulations are performed on smaller arrays of 2690 × 3190 µm$^2$ size using two-dimensional periodic boundary condition (PBC) after discretizing the arrays into cuboidal cells of 5 nm × 5 nm × 15 nm volumes. The lateral cell size is taken smaller than the exchange length of Py (~5.3 nm). The shapes introducing the actual size and rounded corners at the edges have been derived from SEM image of the sample. The material parameters used for simulations are gyromagnetic ratio $\gamma$ = 17.6 MHz Oe$^{-1}$, anisotropy field $H_k$ = 0, saturation magnetization $M_s$ = 860 emu cm$^{-3}$, and exchange stiffness constant $A$ = 1.3 × 10$^{-6}$ erg cm$^{-1}$ [6]. Here, $M_s$ value is taken from our previous work on 15-nm-thick Py film [30].

**V. Results and Discussion:**

We investigate SW dynamics of the BMC sample with varying in-plane bias magnetic field and its orientations using TR-MOKE microscope. The SEM images already showed the lateral dimensions and shapes of the nanodots in the sample as described above. The atomic force microscopy (AFM) image (Fig. 1(b)) shows the topography of the sample and the line scan profiles shown in the inset of Fig. 1(b) show the thickness of the sample to be ~25 ± 3 nm, which demonstrate a slight overshoot of the FIB milling down to the substrate. We observe some depletion near the edges of the dots which may be due to the roughness and deformation at the edges. Time-resolved Kerr rotation data for five different bias fields are presented in Fig. 1(c) after subtracting the bi-exponential background originated from the energy dissipation from electron and spin to lattice system and then from lattice to the surroundings during the remagnetization process followed by the ultrafast demagnetization.

Figure 2(a, b) shows SW spectra obtained from the experimental and simulated time-resolved data after filtering the high frequency noise and performing fast Fourier transformation (FFT) spectra of the time-resolved data, at different bias magnetic field values ($H$). Rich SW spectra are obtained at ϕ = 0° for $H$ = 1080 Oe, which show two separate bands, (mode: 1, 2, 3) and (mode: 4, 5) appearing at higher and lower frequency regimes. The SW modes show strong magnetic field dependence and thus confirm their magnetic origin. Figure 2(c) shows the bias field dependence of SW mode frequencies for experimental and simulated SW modes. The experimental data points corresponding to mode 2 and 3 are well fitted with the Kittel formula [38] and the $M_s$ values obtained are 840 emu.cm$^{-3}$ and 570 emu.cm$^{-3}$, respectively, while the other parameters are found to be similar to the 15-nm-thick Py thin film values. However, the other modes are influenced by the demagnetizing effect from the edges of the dots and affected by the complex interaction hence do not fit well with Kittel formula.

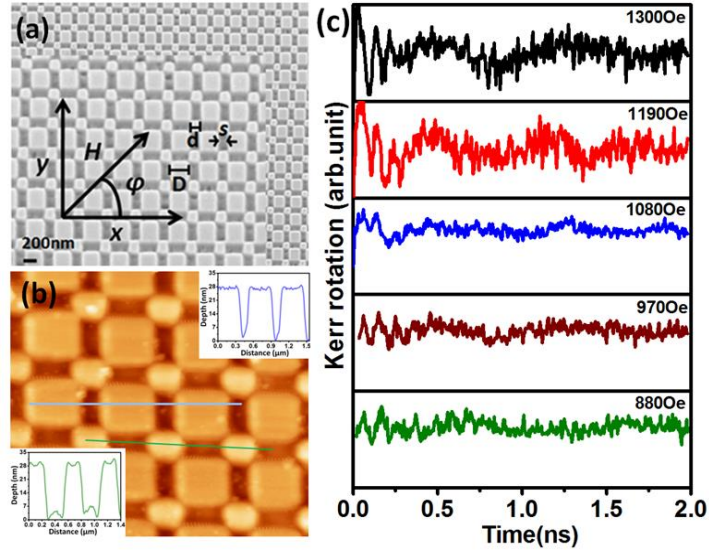

Figure 1: (a) Scanning electron micrograph of 15-nm-thick Py binary magnonic crystal lattice with large dot diameter (D) = 300 ± 5 nm, small dot diameter (d) = 150 ± 5 nm and edge-to-edge separation of the dots (s) = 150 ± 5 nm along with the geometry of the applied magnetic field of the measurement. Here ϕ is the angle between applied bias field and x-coordinate, which was varied from 0° to 180° during the measurement. (b) AFM image and corresponding line scan profile along the dotted line of large dot and small dot are shown in the insets. (c) Background subtracted time-resolved Kerr rotation data for some specific orientations of the bias magnetic field.

The reduction of $M_s$ for mode 3 is due to the collective dynamics of the edge modes of the nanodots, which is localized in strongly demagnetized regions near the edges of the dots [18]. The nominal difference between the simulated and experimental SW mode frequencies may be attributed to the random demagnetized regions at the edges and rounded corners of the dots, which is hard to precisely incorporate in the finite difference method based micromagnetic simulations such as OOMMF as used here.

The SW spectra for different orientations of in-plane bias magnetic field are presented in Fig. 3 (a, b) obtained from experiment and simulation. The modes correspond to different branches of SW bands in the frequency range between 2.5 GHz and 12.5 GHz. The spectral width of these branches narrows down gradually as ϕ varies from 0° to 45°. Thus, at 45°, the frequency band drastically narrows down to 7.5-13.2 GHz. A careful observation shows that the lower frequency modes 4 and 5 get significantly blue shifted (~5 GHz) with the variation of ϕ from 0° to 45°. However, the blue shift in higher frequency modes is not significant enough. With further increase in ϕ, the whole spectra get broadened and the SW branches are red shifted for 45°<ϕ<90°. This trend continues up to 180° in a periodic fashion. Thus, the pronounced effect of the angular dependence is observed for the lower frequency SW modes, and to understand the rotational symmetry of these modes we have plotted their frequencies with ϕ in Fig. 4 (a, b).

The Modes 4 and 5 show clear four-fold rotational symmetry as evidenced from the experimental and simulated results. Modes 1, 2 and 3 show slightly modified four-fold symmetry probably due to the complex structure of the sample and thus ensuing internal field and stray field configurations. The disagreement between experimental and simulated frequencies with variation of ϕ increases in the higher frequency band probably due to the inhomogeneous line broadening of these spectral peaks. This affects the higher frequency modes more due to their smaller variation with angle.

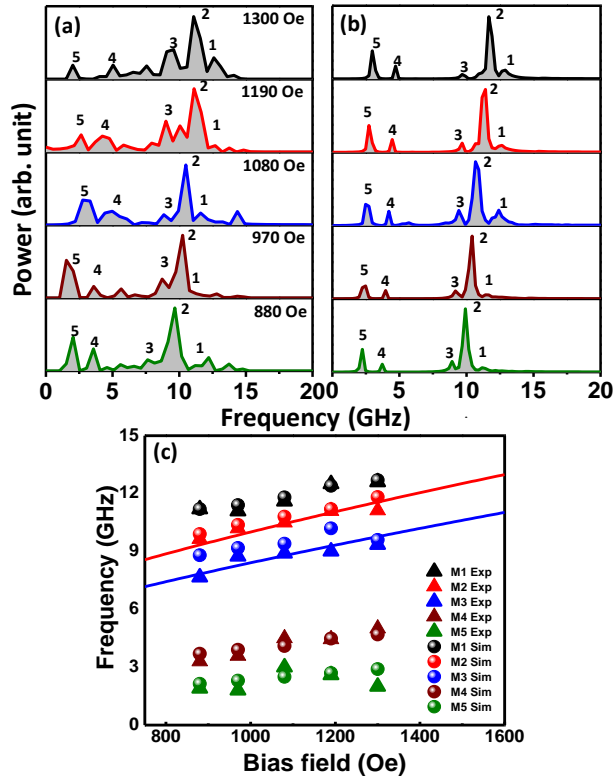

Figure 2: Power vs frequency spectra of the sample with varying strength of bias magnetic field are obtained from using (a) time-resolved MOKE experiment and (b) micromagnetic simulations. (c) Precessional frequencies of different SW modes for (triangular symbols: experimental data, circular symbols: micromagnetic simulation results, solid line: Kittel fit) are plotted as a function of bias field H.

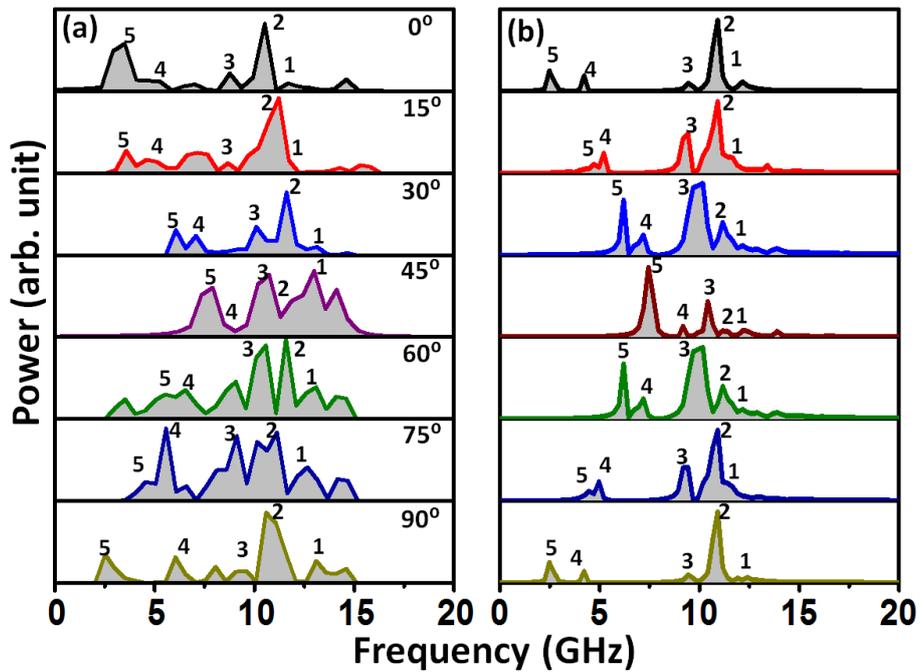

Figure 3: Power vs frequency spectra of the sample with varying orientation of bias magnetic field (ϕ) obtained from using (a) time-resolved MOKE experiment and (b) micromagnetic simulations.

In order to better understand the origin of collective SW dynamics of the BMC, we perform micromagnetic simulations of magnetization dynamics on individual array of large dot (LD) and small dot (SD) separately. We consider again the 4 × 4 elements along with the 2-D PBC and the size of LD and SD same as that of the BMC. Figure 1S (a)-(d) (see supplemental material) shows the comparison of angular variation of simulated SW mode frequencies for the BMC, the arrays of LD and SD. We observe that due to the interaction between the dots, the LDs in the lattice are influenced by the connecting SDs and the nature of SW spectra of BMC gets significantly modified. Some of the modes of LD array are dominating in the spectra for BMC, for example, mode 2 of BMC is mainly originated from the mode 2 of the individual array of LD having a weak four-fold symmetry. On the other hand, no SW modes are observed for individual arrays of LD in the spectra at $f < 8$ GHz whereas strong interaction between the dots in the BMC, edge mode of the SDs induces edge mode of LD with extremely high anisotropic nature (four-fold). This is generally not observed for array of LD alone.

**Power and phase analyses:**

For further understanding of the variation of anisotropic behavior of the SW modes, we simulate the spatial profiles of the SW modes using a home-built code [38]. The simulated power maps for the sample are shown in Fig. 5. At 0°, mode 1 and 2 (12.4 GHz and 10.9 GHz) are the collective modes of the entire array where power is concentrated at the centre of the SDs and LDs respectively. There is very small power in the LDs (SDs) for mode 1 (mode 2), showing, the localization of the modes on specific dot size. Modes 3 and 4 (9.4 and 4.2 GHz) are the modified edge mode with Damon-Eshbach (DE) nature of the LDs and pure edge mode for SDs, respectively. Mode 5 is the collective edge mode of both the dots in the array, due to strong dipole-exchange interaction at the connected corners. This frequency-selective collective excitation in the LDs (SDs) may find interesting applications in magnonic devices. On the other hand, the collective excitation in both dots may be further explored for efficient SW propagation as shown later in this article.

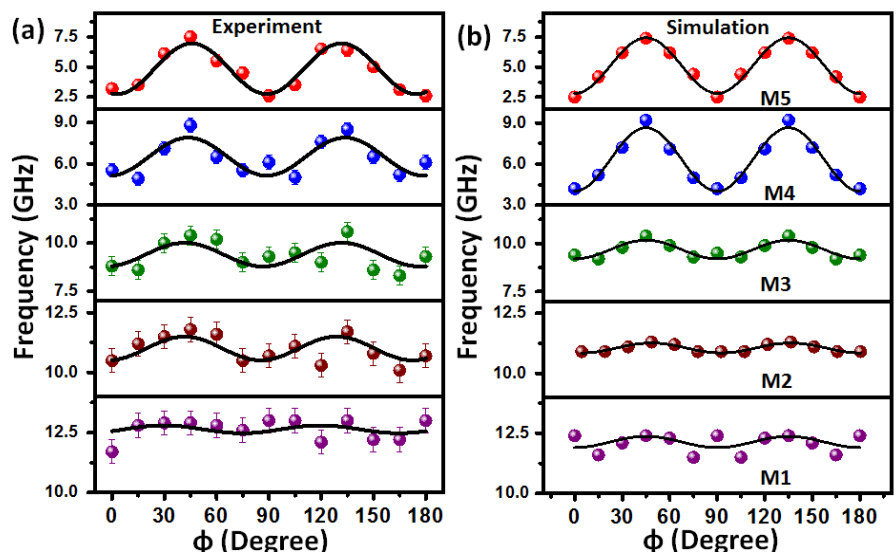

Figure 4(a): Angular variation of different spin-wave modes of the (a) experimental and (b) simulated sample are shown for different spin-wave modes. Those are marked as M1, M2. M3. M4 and M5 inside the figure. The data points are shown by symbols and the black solid lines represent the theoretical fitting.

Further, at 15°, we observe that the frequency selective nature of the collective SW excitation disappears and the dots start to interact leading towards extended modes in the array. Mode 1 shows quantized nature in LDs and SDs with quantization number (n = 7; n` = 3), respectively. Mode 2 is no more centre mode of the large dot, but becomes quantized mode both in the LD and SD with quantization number, n = 5 and n` = 3, respectively. Mode 3, which was the edge mode of LD at 0°, becomes quantized mode at 15° with n = 3. Modes 4 and 5 are interacting edge modes of both the LD and SD of the array. On further rotation of the bias field to 30°, the interaction between the dots grows and can be clearly seen from the power profile. Modes 1, 2 and 3 show the mode quantization (n = 7, 7, 7 and n` = 3, 5, 5). Modes 4 and 5 remain interactive edge modes. At ϕ = 45°, the interaction between the large and small dots gets maximized and all the modes show the extended nature. A mode conversion is observed where power of the SWs start to channelize diagonally through both LDs and SDs forming extended SW mode. Modes 1, 2, and 3 are transverse quantized modes with higher quantization numbers (n = 11, 9, 9; n` = 5, 3, 3), respectively. On further increase of ϕ, the SW modes slowly lose the extended nature and become fully localized as ϕ approaches 90°.

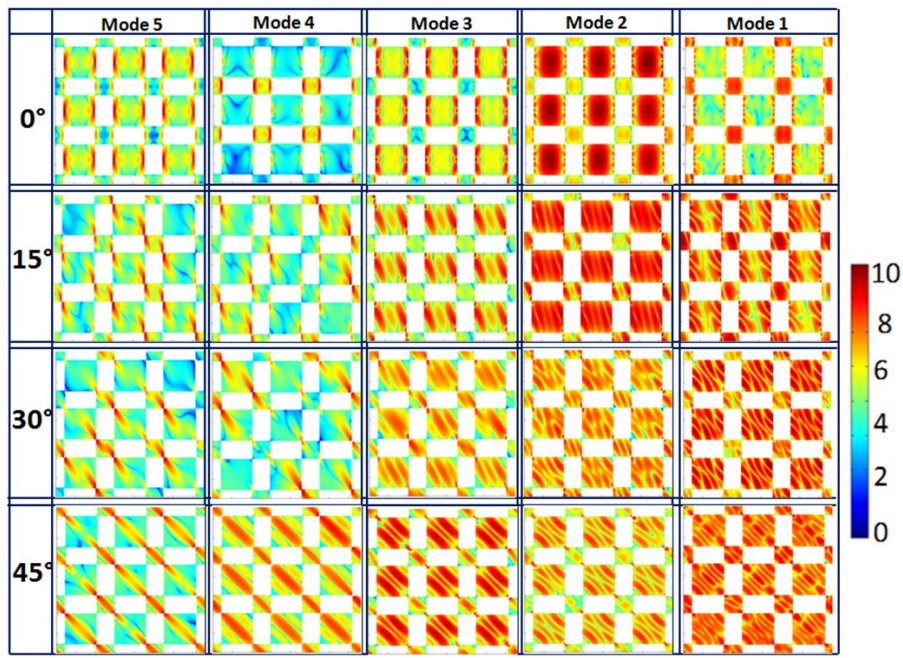

Figure 5: Power profile for the spin-wave modes of the sample for different orientations of the in-plane bias magnetic field. The color bar is shown on the right side of the figure and is represented in dB.

After comparing with the power profile for individual arrays of LDs and SDs (see Fig. 2S in the supplemental material) we understand the nature of SW modes in the BMC with more clarity. Frequency selective localization of SW modes observed in BMC (mode 1 and 2) appears due to the centre and edge modes of arrays of LDs and SDs. The modified edge mode of the LD is dominating at $\phi = 0°$ for BMC. With increasing $\phi$, the nature of the high frequency modes becomes quantized and these modes are responsible to form the propagating modes in BMC, in presence of SDs as connector. Interestingly, the edge mode of array with SD causes appearance of pure edge modes in the LD when they are connected via dipole-exchange coupling.

**Local excitation of spin-wave modes:**
To investigate the nature of propagation of SWs in the sample with the orientation of the bias magnetic field, we locally excite SWs at the centre of the array over a rectangular area of 300 × 100 nm² for two values of $\phi$ of 0° and 45°. For SW excitation, we use a time-varying field of "sinc" profile as given below with field amplitude of $H = 50$ Oe, where a frequency cut-off of 25 GHz was chosen, which is well above the frequencies of the observed modes.

$$H_{ext} = H\left(\frac{\sin y}{y}\right) \quad (1)$$

The power profiles of the SW modes were further calculated from the locally excited dynamics using method as described above. Figure 6 shows the spatial distribution of power of SW modes 1, 2 and 5 for 0° and 45° angles, respectively. These modes show conversion from localized to propagating nature with the variation of $\phi$. For mode 1, power is confined at the point of excitation for 0°. However, at 45°, the collective mode propagates uniformly over a large area of the array in the DE geometry. For mode 2, at $\phi = 0°$ the uniform mode of the LD propagates in the DE geometry mediated by the edge mode of the SD. On the contrary, the propagation of mode 2 decays very fast for ϕ = 45°. Mode 5, on the other hand, shows little or no propagation of the edge mode of the LD at ϕ = 0°. However, at ϕ = 45°, a continuous propagation channel opens up through the LD and SD leading towards strong propagation of the SW over a long enough distance. The characteristics of other SW modes after local excitation is described in the Fig. 3S of the supplemental material. Thus, we demonstrate a strong variation of propagation of various SW modes in this BMC simply by varying ϕ, which is an external parameter. These results can open up new possibilities for construction of SW filter with a broad band distribution as well as directional SW coupler.

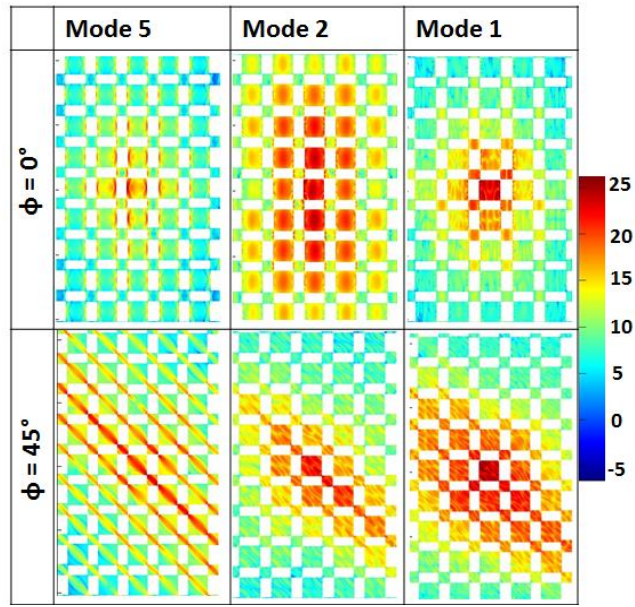

Figure 6: Propagation of mode 1, 2 and 5 of BMC at two different orientation of bias field after local excitation at the middle of the array. The color scale is represented in dB.

**Magnetostatic field analysis:**

The physical mechanism underlying the occurrence of various SW modes with the change in ϕ can be understood from the simulation of distribution of magnetostatic field lines within the array by varying ϕ using LLG micromagnetic simulator [39]. Figure 7(a) shows contour maps of magnetostatic field distributions of the array for specific orientations of the bias magnetic field, where three important regions along the y-direction are taken into account: region A: across the interfaces of LD and SD (yellow dotted line); region B: across SD (green line) and region C: across LD (blue line). We plot the above fields taken within the boxed region in Fig. 7(b) – (d) for ϕ = 0, 15, 30 and 45°. These clearly show non-uniform distribution of the magnetostatic fields across the lines A, B and C.

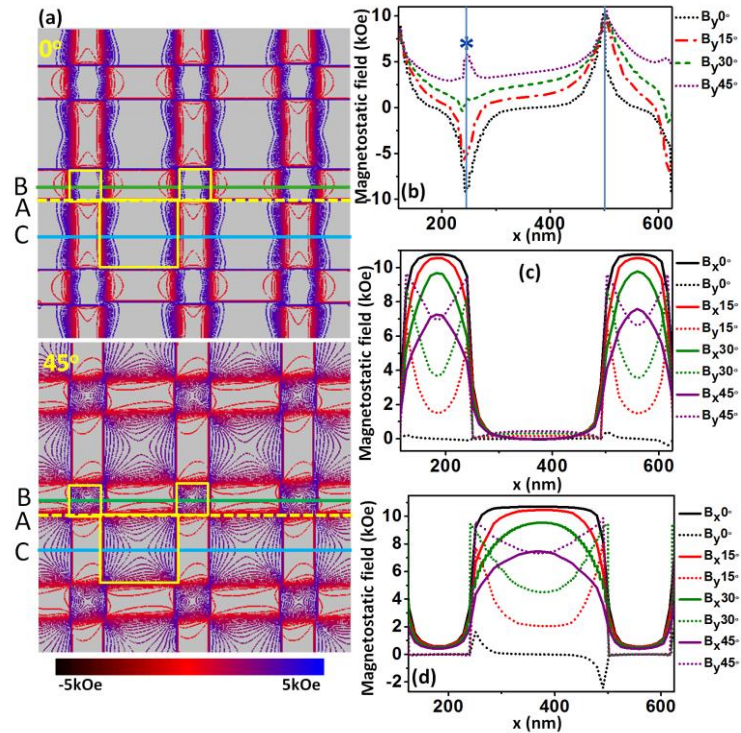

Figure 7: (a) Contour maps of simulated magnetostatic field distributions at ϕ = 0° and 45° respectively. The color scale is given on the right side. The magnetostatic fields distributions for (b) from middle of the both dots (dotted yellow line), (c) small dots (green line) and (d) large dot (blue line) respectively at different orientation of the bias magnetic field is taken and are shown for the dotted box of the sample regions.

From the plot (Fig. 7 (b)), it is clear that the magnitude and sign of maxima and minima of the magnetostatic field at the two edges of the large and small dots depend strongly on variation of ϕ. The maxima and minima occur primarily due to the modulation of the magnetic field at the overlap region of the two dots at the corners. Notably, at 0°, the magnetostatic field has its maximum (8.83 kOe) and minimum (-9.12 kOe) value at the left and right edges of the dot, respectively (see Fig. 7 (b)). Furthermore, on comparing the profiles of the magnetostatic fields for ϕ = 0°, 15°, 30° and 45°, we observe that the minima convert into maxima (at the region marked with '*') with the drastic variation in the magnetostatic field from -9.1 kOe to 5.7 kOe. This is due to the variation of the internal spin configuration within the dots with ϕ. This explains the increase in frequency of the two lower frequency modes (mode 4 & 5) with the increase in ϕ (see Fig. 3 (a) and (b)).

Detailed inspection of the $B_x$ and $B_y$ components of the magnetostatic field profiles in SD and LD shown in Fig. 7 (c) and (d) shows the relative change of the magnetostatic field with the variation of ϕ. At ϕ = 0°, the values of $B_x$ in regions B and C are 10.78 kOe and 10.70 kOe, while the value of $B_y$ is negligibly small in both regions (at the centre of the lLDS and SDs), which are responsible for the centre mode of small and large dots (see Fig. 5). At 15°, the values of $B_x$ ($B_y$) reduce (increase), 10.57 kOe (1.53 kOe) in region B and 10.54 kOe (2.05 kOe) in region C. At 45°, we observe an overlap of $B_x$ and $B_y$ with values of 7.26 kOe and 6.95 kOe in region B and 7.45 kOe and 7.32 kOe in region C, respectively. Thus, the variation of $B_x$ and $B_y$ at different bias field orientations appears to be responsible for the variation of the SW modes, including conversion from localized to propagating nature. The comparison of stray field distribution for individual arrays of large and small dots at the ground state under applied bias field, is shown in Fig. 4S of the supplemental material. We observe that for the array with LD and SD, the magnetostatic coupling is much weaker and the dots retain their individual behavior. The amount of contribution from the magnetostatic coupling with the neighboring dots is significantly high for large and small dots when they are connected together and thus confirms the strong coupling of the SW modes in BMC.

## VI. Conclusion:

We fabricate a new type of binary magnonic crystal where two square Py nano-dots are connected diagonally using FIB lithography. The magnetization dynamics of the sample is measured by using time-resolved magneto-optical Kerr effect microscope with varying magnitude and in-plane orientation (ϕ) of the bias magnetic field. At ϕ = 0°, a frequency selective collective localization of SW modes is observed within the array. We observe band narrowing due to the conversion of localized mode to extended mode on rotation of the bias field orientation within 0°≤ ϕ ≤ 45°. Upon further rotation of ϕ above 45°, the SW modes gradually loose the extended nature and become fully localized at 90°. All our results agree well with micromagnetic simulations. The magnetostatic stray field analyses permit us to quantitatively explain the observed conversion of SW modes with varying in-plane bias magnetic field orientations. This is due to the variation in coupling strength of magnetostatic field lines of large and small dots at different bias field orientations. The mutual coupling of magnetic field lines shows strong inhomogeneity at the interfaces of the sample (large and small dots). We also numerically show that the sample provide control over SW propagation length in the lattice with the orientation of the magnetic field. This new type of binary magnonic crystal will open up new possibilities for construction of various spintronic and magnonic devices.

**Acknowledgments:**
The authors acknowledge the SGDRI (UPM) project of IIT Kharagpur for support. AB acknowledges S. N. Bose National Centre for Basic Sciences for financial support (Project No. SNB/AB/18-19/211). SM and KD acknowledge DST for INSPIRE fellowship, JS acknowledges DST for Ramanujan fellowship, while SC acknowledges S. N. Bose National Centre for Basic Sciences for senior research fellowship.

**Supplemental Information**

1. **Anisotropic nature of spin-wave modes for individual arrays of large and small dots:**

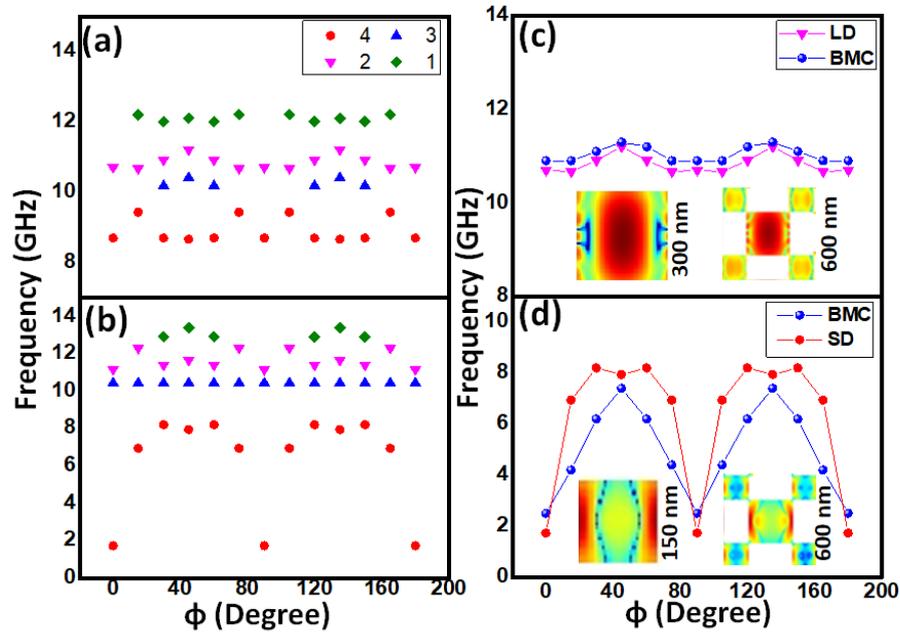

Figure 1S: Angular variation of different spin-wave modes corresponding to the array of (a) larger dot (LD) and (b) smaller dot (SD). Mode numbers are indicated in numeric figures. Comparison between angular dispersion for (c) mode 2 of BMC and mode 2 of an array of LD, (d) mode 5 of BMC and mode 4 of an array of SD. The inset shows the corresponding power profile of the modes.

The basis of binary magnonic crystal (BMC) consists of a large dot (LD) of 300 nm width connected diagonally with small dot (SD) of 150 nm width. This binary combination gives rise to complicated exchange and dipolar coupling in the system and thus the spin-wave (SW) spectra experience mode conversion and band narrowing, with the variation of in-plane orientation (ϕ) of bias magnetic field. To understand the origin of the rich anisotropic behavior of the SW modes, we have simulated the magnetization dynamics of individual arrays of LD and SD with similar size and inter-dot separation to the experimental sample. In the micromagnetic simulation, we consider the 4 × 4 elements along with the two-dimensional (2-D) periodic boundary condition (PBC) and all the material parameters are used as described previously in the simulation details of this article. Figure 1S (a) shows the variation of frequencies for all the SW modes obtained for an array with LDs. The modes show anisotropic behavior and possess rotational symmetry by varying in a periodic fashion at an interval of 90°. The SW modes for an array with SDs also show significant dispersion with ϕ (see Fig. 1S (b)). The mode 4 at lowest frequency regime appears with strong four-fold anisotropy. Comparison of the nature of modes in Fig. 1S (c) shows that mode 2 of BMC (Fig. 4) is mainly originated from the mode 2 of the individual array of LD having a weak four-fold symmetry. These modes are centre mode of the LD distributed uniformly through the whole array and power profiles of those are shown in the inset of Fig. 1S (C). Surprisingly there is no SW modes observed for individual arrays of LD in the spectra at $f < 8$ GHz. However, the strong interaction between the dots in the BMC, edge mode of the SD induces edge mode of LD with extremely high anisotropic nature which is generally not observed for isolated dot arrays alone.

For LD array, two distinct modes appear in the SW spectra which are centre mode and modified edge mode of the dots (see Fig. 2S) distributed uniformly. With increasing ϕ, a mode conversion is observed and the nature of the modes becomes quantized with quantization number, n > 3. The low-frequency mode shows standing SW-like nature at ϕ = 45°. Though no prominent edge mode is found for this array. For SD array, three modes appear with isolated nature, *i.e.* intrinsic edge mode, modified edge mode and centre mode of the individual dots. Similar to the LD array, here also we observed the mode quantization of the high-frequency modes and that is because of the shape anisotropy present in the dots.

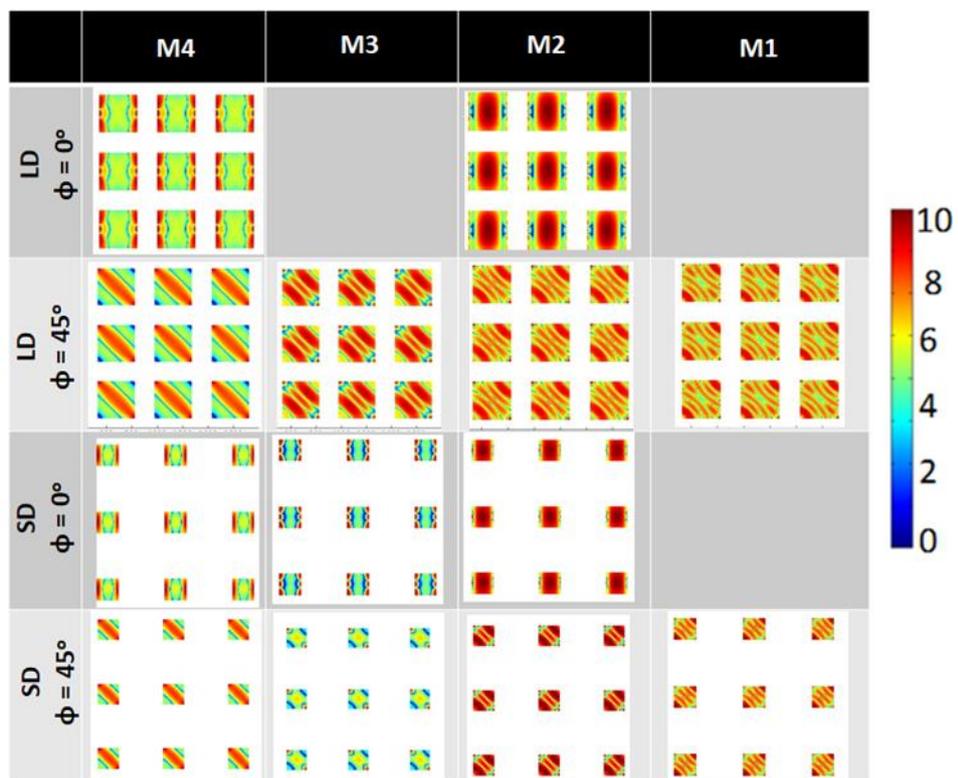

Figure 2S: Power profile for the spin-wave modes from arrays of LD and SD at ϕ = 0° and 45°, respectively. The color bar is shown on the right side of the figure and is represented in dB

2. Spin-wave propagation due to local excitation in the BMC: Numerical simulation

We observe that different SW modes of BMC propagate in a different manner according to their spatial characteristics in the array and availability of channel for propagation. Modes 3 and 4 of BMC (shown in Fig. 3(S)) are mainly dominated by the edge modes of LD and SD, respectively. At ϕ = 0°, the pinning at the respective edges, does not allow the modes to distribute power in the neighborhood, whereas with increasing angle, the possibility of opening up of the new channels through the exchange coupled regimes of the dots increases. Thus, those quantized modes can propagate with reasonable intensity across the diagonally extended channels in the array.

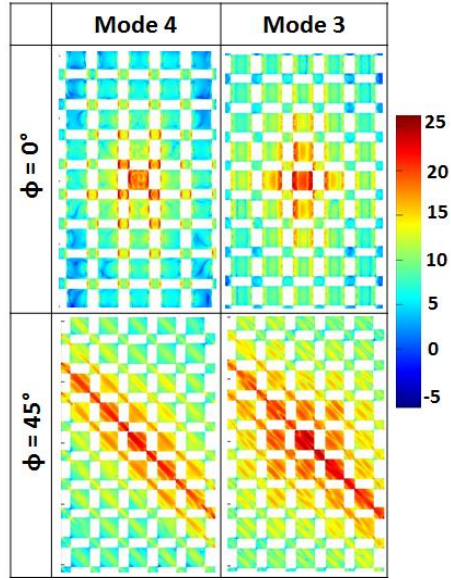

Figure 3S: Propagation of modes 3 and 4 of BMC at two different orientations of bias magnetic field after local excitation at the middle of the array. The color scale is represented in dB.

3. Analysis of magnetostatic coupling in the individual arrays of larger and smaller dots:

To have a deeper understanding of the dipolar interaction interferring the mode profile of crystals we have simulated the magnetostatic field distribution using LLG micromagnetic simulator. The magnetic charges at the edges of the LDs (see Fig. 4S (a)) participate in the dipolar coupling with the neighboring dots at ϕ = 0° and 45°. With increasing angle, the interactions become more complicated. But for arrays of SDs, the magnetostatic coupling is much weaker and the dots retain their individual behavior. Figure 4S (b) shows the line scan of the stray field distribution across the dots in the individual arrays as well as in BMC and the corresponding values are shown in Table 1 for the comparison. The amount of contribution from the magnetostatic coupling with the neighboring dots is different for LD and SD when they are connected together.

Table 1: Values of magnetostatic fields for BMC, LD and SD at ϕ = 0° and 45°

| ϕ (Degree) | Values of magnetostatic field (kOe) | | | |
| --- | --- | --- | --- | --- |
| | LD | LD of BMC | SD | SD of BMC |
| 0 | 9.93 | 10.7 | 9.70 | 10.78 |
| 45 | 7.40 | 7.45 | 7.04 | 7.26 |

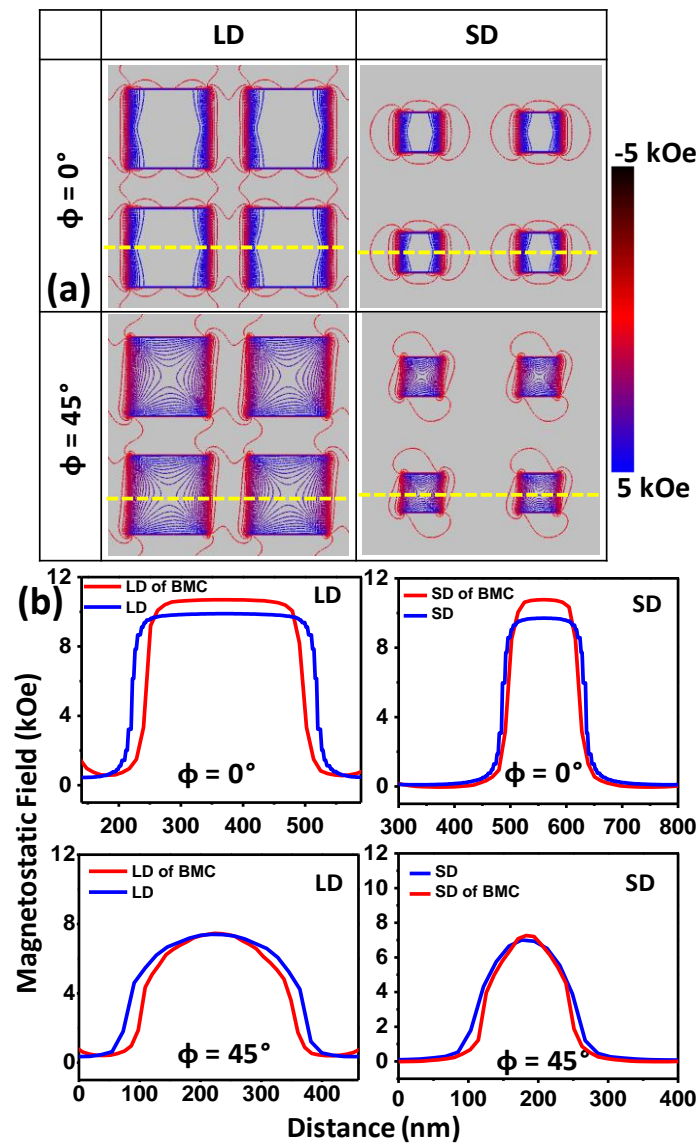

Figure 4S: (a) Magnetostatic field distribution for the arrays with LD and SD at $\phi = 0°$ and $45°$, respectively. (b) The variation of x-component of the field values obtained from the line scan in respective orientation is presented. The red and blue solid lines represent the value of stray fields obtained from BMC and individual arrays, respectively.